# Kinetic Effects in Strong Langmuir Turbulence

AKIMOTO, Kazuhiro  HOJO, Hitoshi [1)] and SAITOU, Yoshifumi [2)]

*School of Science and Engineering, Teikyo University, Utsunomiya, Japan*

1) *Plasma Research Center, University of Tsukuba, Tsukuba, Japan*

2) *School of Engineering, Utsunomiya University, Utsunomiya, Japan*

*E-mail: akimoto@ees.teikyo-u.ac.jp*

**Abstract**

Kinetic effects with regard to a one dimensional Langmuir soliton-like pulse are investigated. Though thus far mainly transit-time accelerations have been investigated regarding strong Langmuir turbulence, it is found that ponderomotive reflections (generalized nonlinear Landau damping) may play important roles also. The former may diffuse fast electrons up to relativistic energies, while the latter reflects slow electrons as well as ions that have speeds comparable with the group velocity of the pulse, and tend to form flat-top electron distributions at and around the quasi-soliton.

## 1. Introduction

Accelerations of charged particles by Langmuir solitons have been studied extensively[1] as the foremost dissipation mechanisms in strong Langmuir turbulence. Here, particle accelerations by a one-dimensional Langmuir soliton-like pulse are investigated in detail. When relatively long, Langmuir waves have large enough amplitudes, they are subject to modulational instabilities[1]; the modulational instabilities are saturated by forming Langmuir solitons that are supposedly stable, say, in a sufficiently intense magnetic field. Such a 1D Langmuir soliton field $E(z, t)$ as well as the corresponding density perturbation $N(z, t)$ are given as functions of the space $z$ and time $t$ in a normalized form respectively as follows[1,2]:

$$E(z,t) = E_0 \{\cosh[K_0(z - v_g t)]\}^{-1} \exp[i(K_1 z - \Omega t)] \qquad (1)$$

$$N(z,t) = -6K_0^2 \{\cosh[K_0(z - v_g t)]\}^{-2} \qquad (2)$$

Here, $E_0^2 = 24K_0^2[1 - 9K_1^2(m_i/m_e)]$, $v_g = 3K_1$ is the group velocity of the soliton, and $\Omega = 1 + 3(K_1^2 - K_0^2)$. The time is normalized by the electron plasma frequency as $\omega_e t$, the space by the Debye length as $z/\lambda_e$, the velocity by the electron thermal velocity as $v/v_e$, and finally the electric filed as $eE/(m_e\omega_e v_e) = \varepsilon_0 E/(4n_e k_B T_e)^{1/2}$, respectively; other notations are standard. It can be proved that 1D Langmuir solitons are stable in the absence of dissipation,[1] which is not included in Eqs.(1) and (2). However, the dissipation of a pulse cannot actually be ignored if there exit efficient mechanisms for particle accelerations, which are the subject of the present study. In fact, C.H. Ling et al.[2] have performed 1D Vlasov simulations of various Langmuir solitons, and showed that a high-energy tail is formed on the electron distribution function as well as that initially propagating Langmuir soliton may be decelerated rather efficiently unless $T_e \gg T_i$. The former is well known, and may be explained via transit-time acceleration[1] [TTA], but the latter, which the authors attributed to nonlinear Landau damping, requires a more explicit explanation. The result of the simulation by C.H.Ling et al.[2] is rather surprising because 1D Langmuir solitons have been believed stable. In the present study, however, to keep coherence with the previous studies of particle accelerations by a Gaussian pulse, a Gaussian soliton-like pulse is adopted as below.

$$E(z,t) = E_0 \exp\left[-\{(z - v_g t)/L\}^2\right] \cos[i(\omega_e t + \theta)] \qquad (3)$$

Here, the parameter $L$ determines the pulse size, $\omega_e$ is the plasma frequency, and $\theta$ is a phase constant. With this form of pulse, various acceleration mechanisms have been identified previously, and their theoretical estimates are readily available.[3]

To investigate particle acceleration analytically or numerically the one-dimensional relativistic equation of motion is used. It was previously found that a pulse of the form (3) may accelerate/decelerate charged particles through two mechanisms, i.e., TTA and ponderomotive reflections[PR].[3] TTA, after penetrating a pulse, yields following velocity shifts:

$$\Delta v = \frac{qE_0 \cos\theta \Delta t}{m\gamma_0} \exp\left\{-\left(\frac{\omega_e \Delta t}{2}\right)^2\right\}, \qquad (4)$$

where $\Delta t = L/(v_g - v_0)$ is a type of transit-time.

When $v_0$ is close to $v_g$, the transit-time is extremely large, and thus the velocity shift (4) becomes negligible. When $\omega_e \Delta t = \sqrt{2}$ (being close to $\pi$), the velocity shift (4) is maximized. The maximum velocity shift may be calculated easily from (4). As the transit-time $\Delta t$ is made longer the velocity shift decreases. For high velocity particles the transit-time $\Delta t$ is reduced, and the velocity shift becomes proportional to $\Delta t$ or inversely proportional to $v_0$. Only in the limit of $v_0 = c$ the maximum velocity shifts by TTA equals zero. Namely, if sufficient wave energy and interaction time are available, TTA may accelerate particles to relativistic energies.

On the other hand, PR is effective for particles resonant with $v_g$, i.e., $v_g + v_{ref} > v_0 > v_g - v_{ref}$ with $v_{ref} = qE_0/(\sqrt{2}m\omega_e)$; the maximum quiver velocity divided by $\sqrt{2}$, and the resultant velocity shift is $\Delta v = 2(v_g - v_0)$.

## 2. Results
### 2.1 Transit-time acceleration and nonlinear Landau damping

In this section, the equation of motion will be solved numerically, and the resultant velocity shifts will be compared with theoretical results. As one of our objectives is to elaborate on the simulation results of C.H. Ling et al.,[2] we will set physical parameters as their simulations, i.e., the ion-to-electron mass ratio is 100, $c/v_e=100$, and $K_0=3/15$. The last one leads to $E_0=0.00979$. for electrons interacting with a standing soliton. $L_n = L/(c/\omega_e) = 0.15$. In Fig.1 we first note that TTA in this case is effective for fast electrons with $v_0 > 4 v_e$. Moreover, numerical results are in excellent agreement with analytical ones (4). In Fig.1 and Eq.(4), the possibility is shown that if sufficiently long interaction time is allowed, TTA may drive electrons to relativistic energies. Thus, it is quite understandable that in the simulations of C.H. Ling et al.[2] among others[1] tails were formed on background electron distribution functions. TTA may decelerate some fast electrons to lower velocities. Hence, as is well known, TTA diffuses electrons in velocity space. Meanwhile, slow electrons with $v_0 \leq 0.6\ v_e$ are subject to PR, through which electrons resonant with the soliton, i.e., $v_0 \approx v_g = 0$ are reflected. The numerical results are in excellent agreement with the theory. Though in the velocity space the active region of TTA is much broader than that of PR, the number of particles subject to PR exceeds that to TTA. Hence, PR could become more important than TTA.

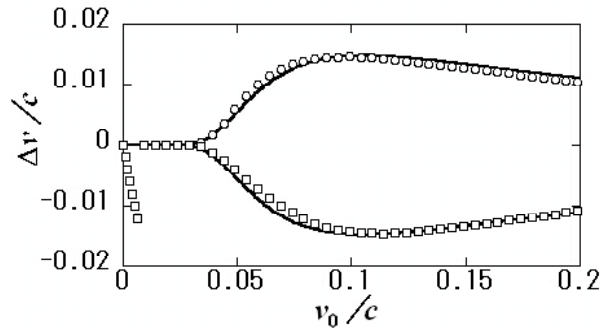

Fig.1 Electron velocity shifts as a function of $v_0$. Here, the maximum (circles) as well as minimum (squares) velocity shifts after the interaction with a Langmuir soliton-like pulse (3) are plotted. The initially low velocity electrons are reflected by PRs, while high velocity ones are subject to TTA. TTA is not so effective for heating ambient electrons.

**2.2 Effects on distribution functions**

Here, we solve Eq.(1) for distributions of particles to see the total effects of particle accelerations. A simulation box of $300\lambda_e$ with $\lambda_e$ being the Debye length with a periodic boundary condition is adopted. As many as 15,520 electrons are used. This way, it becomes possible to evaluate the importance of the two competing acceleration mechanisms. $v_e/c=0.01$ is assumed. Fig.2 shows the electron distribution functions at and around the soliton ($120\lambda_e < z < 180\lambda_e$), which is located at the center of the box, plotted as a function of time. As time elapses, low energy electrons are reduced. This is due to the fact that low energy electrons are reflected by the Langmuir pulse via PR. Because of PR electron density is depleted where the soliton is present.

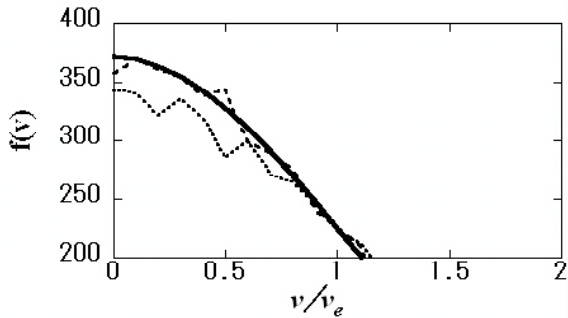

Fig.2 Electron distribution functions inside the Langmuir soliton plotted at various times. The solid, dotted and broken curves correspond to $\omega_e t = 0$, 17.3($5.5\pi$), 18.8($6\pi$), respectively.

We separately investigate the effects of TTA on an electron beam. An electron beam with $v_0=0.05c$ is made to interact with a soliton. It is found that at $\omega_e t = 100$ a tail is formed on the high-energy side of the beam and the fastest electrons gained the velocities slightly less than $11v_e$, while the low-energy side remains essentially Maxwellian. This trend becomes more conspicuous at $\omega_e t = 200$. These results can be explained by TTA expressed in Fig.1. Beam density also is depleted somewhat at the position of the soliton.(not shown) Finally, let us point out that in the presence of solitons beam particles tend to be accelerated to form an exponential energetic tail. This is owing to the fact that the beam-plasma instabilities generate waves, the phase velocities of which are close to the beam velocity. This indicates that the beam generated turbulence is first dissipated by beam particles rather than the ambient particles as usually thought.

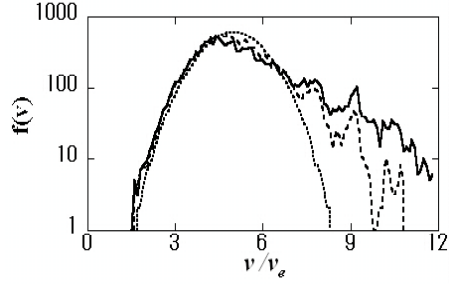

Fig.3 Evolution of a beam distribution function interacting with a Langmuir soliton. The dotted curve shows the initial distribution function, whereas the broken and solid curves show the ones at $\omega_e t = 100$ and 200, respectively. Only the high-energy side of the distribution function is essentially developing to form almost an exponential energetic tail.

## 3. Conclusions

Present analyses of particle acceleration mechanisms with regard to Langmuir solitons reveal that for electrons PR (nonlinear Landau damping) also may become important. It is found that PR is responsible for the formation of non-Maxwellian electron distribution functions with reduced low-energy electrons at and around Langmuir solitons. Finally, in strong Langmuir turbulence beam particles may be accelerated, and form an exponential energetic tail. It may become the most significant dissipation source.

**References**
[1]. P. A. Robinson, Rev. Mod. Phys. **69**, 507(1997).
[2]. C.H. Ling, J.K.Chao, and C.Z.Cheng, Phys. Plasmas, **2**, 4195(1995).
[3]. K. Akimoto, *Phys. Plasmas*, **4**, 3101 (1997); ibid. **9**, 3721(2002); ibid. **10**, 4224(2003).